\documentclass[a4paper,twocolumn,11pt,accepted=2025-05-08]{quantumarticle}
\pdfoutput=1
\usepackage[utf8]{inputenc}
\usepackage[english]{babel}
\usepackage[T1]{fontenc}
\usepackage{amsmath}
\usepackage{hyperref}

\usepackage{tikz}
\usepackage{lipsum}

\usepackage{setspace} %to set 1.5 spacing in text
\usepackage{graphicx}
\usepackage{amsmath}
\usepackage{color}
\usepackage{amsmath}
\usepackage{mathtools}
\usepackage{amssymb}
\usepackage{verbatim}
\usepackage{physics}
\usepackage{latexsym}
\usepackage{enumerate} % to change the numbering of 'enumerate'
\usepackage{bm} % for bold face mathematical symbols
\usepackage[caption=false]{subfig}
\usepackage{multirow}
\usepackage{booktabs}
\usepackage{siunitx}
\usepackage{bm,bbm}

\usepackage{float}

\usepackage{lipsum}

\begin{document}

\title{Classical Benchmarks for Variational Quantum Eigensolver Simulations of the Hubbard Model}

\author{Antonios M. Alvertis}
\email{antonios.alvertis@austin.utexas.edu}
\affiliation{KBR, Inc., NASA Ames Research Center, Moffett Field, CA 94035, United States}
\affiliation{Department of Physics, The University of Texas at Austin, Austin, TX 78712}
\affiliation{Oden Institute for Computational Engineering and Sciences, The University of Texas at Austin, Austin, TX 78712}
\author{Abid Khan}
\affiliation{Department of Physics, University of Illinois Urbana-Champaign, Urbana, IL, United States 61801}
\author{Thomas Iadecola}
\affiliation{Department of Physics and Astronomy, Iowa State University, Ames, IA 50011, USA}
\affiliation{Ames National Laboratory, Ames, IA 50011, USA}
\author{Peter P. Orth}
\affiliation{Department of Physics and Astronomy, Iowa State University, Ames, IA 50011, USA}
\affiliation{Ames National Laboratory, Ames, IA 50011, USA}
\affiliation{Department of Physics, Saarland University, 66123 Saarbr\"{u}cken, Germany}
\author{Norm M. Tubman}
\email{norman.m.tubman@nasa.gov}
\affiliation{NASA Ames Research Center, Moffett Field, CA 94035, United States}

\maketitle

\begin{abstract}
  Simulating the Hubbard model is of great interest to a wide range of applications within condensed matter physics, however its solution on classical computers remains challenging in dimensions larger than one. The relative simplicity
of this model, embodied by the sparseness of the Hamiltonian matrix, allows for its efficient
implementation on quantum computers, and for its approximate solution using variational algorithms such as the
variational quantum eigensolver. While these algorithms have been shown to
reproduce the qualitative features of the Hubbard model, their quantitative
accuracy in terms of producing
true ground state energies and other properties, and the
dependence of this accuracy on the system
size and interaction strength, the choice of variational 
ansatz, and the degree of spatial inhomogeneity in the model, remains unknown. Here we present a 
rigorous classical benchmarking study, demonstrating
the potential impact of these factors on the accuracy
of the variational solution of the Hubbard
model on quantum hardware, for systems with up to $32$ qubits. We find that even when using the most accurate wavefunction ans\"{a}tze for  the Hubbard model,
the error in its ground state energy and wavefunction plateaus for larger lattices, while stronger electronic correlations magnify this issue. Concurrently, 
spatially inhomogeneous parameters and the presence of off-site Coulomb interactions only have a small effect on the accuracy of the computed ground state energies.  Our study highlights the capabilities and limitations of current approaches for solving the Hubbard model on quantum hardware, and we discuss potential future avenues of research. 
\end{abstract}

\section{Introduction}

Understanding strongly-correlated electron
systems is one of the central
challenges of condensed matter physics. 
Materials with strong electronic interactions exhibit a wealth of interesting properties, including superconductivity~\cite{doi:10.1126/science.1071122,Li2019}, Mott insulating
behavior~\cite{doi:10.1126/science.aam9189,Grytsiuk2024}, excitonic ground states~\cite{Jia2022,Ma2021}, non-Fermi liquid behavior~\cite{RevModPhys.73.797}, competing and intertwined orders~\cite{doi:10.1126/science.1107559,RevModPhys.87.457,RevModPhys.75.913}, and non-trivial magnetism~\cite{PhysRevB.93.054429,Guo2018}. One of the most famous models that is
capable of capturing such phenomena is the Fermi-Hubbard model~\cite{arovasHubbardModel2022,doi:10.1126/science.aal5304,PhysRevLett.110.216405,PhysRevB.75.193103,PhysRevB.89.035139,PhysRevLett.62.591,PhysRevB.85.165135,PhysRevLett.120.247602,RevModPhys.68.13,PhysRevLett.70.3143,arute2020observation,tubman2018postponing}. In its canonical form, its Hamiltonian is written as
\begin{align}
    \label{eq:basic_Hubbard}H= -t\sum_{\sigma}\sum_{\langle \mathbf{R}\mathbf{R}'\rangle}a_{\mathbf{R}}^{\sigma \dagger}a_{\mathbf{R}'}^{\sigma}+ U\sum_{ \mathbf{R}}n_{\mathbf{R}\uparrow}n_{\mathbf{R}\downarrow},
\end{align}
where $\mathbf{R}$ is the lattice vector
corresponding to a site within a
lattice, $\sigma$ the electron spin, $t$
a hopping integral between nearest neighbors ($\langle \cdot \rangle$ denotes nearest-neighbor summation only), $U$ a repulsive on-site interaction, $a_{\mathbf{R}}^{\sigma \dagger}$ ($a_{\mathbf{R}}^{\sigma}$) an
operator creating (destroying) an electron
of spin $\sigma$ at site $\mathbf{R}$, and
$n_{\mathbf{R}\uparrow}=a_{\mathbf{R}}^{\uparrow \dagger}a_{\mathbf{R}}^{\uparrow}$ the number operator for spin-up electrons at site $\mathbf{R}$, and similar for spin-down.

While the Hubbard model can qualitatively capture
the behavior of strongly-correlated systems,
it does not generally offer quantitative predictive accuracy of observables. However, in recent years
several approaches have been put forward
in order to derive material-specific Hubbard models with additional complexity to
that of Eq.\,\eqref{eq:basic_Hubbard}, including \emph{ab initio} downfolding~\cite{Nakamura2021,Zheng2018,PhysRevB.92.054515}, embedding techniques~\cite{Vorwerk2022,PhysRevX.11.021006} and more. The general framework of these schemes entails  electronic structure calculations
within, \emph{e.g.}, density functional theory to derive the parameters entering
the Hubbard model for a given material, 
allowing also for multiple orbitals per site, spatial anisotropy, long-range 
interactions, \emph{etc.}, as a means of
capturing the complexity of real materials~\cite{aryasetiawanFrequencydependentLocalInteractions2004, kentPredictiveTheoryCorrelated2018}.
For example a recently proposed approach
to \emph{ab initio} downfolding generates
the following Hamiltonian for a material~\cite{Nakamura2021}:
\begin{align}
    \label{eq:Hamiltonian}
    H&= -\sum_{\sigma}\sum_{\mathbf{R}\mathbf{R}'}t_{\mathbf{R}\mathbf{R}'}a_{\mathbf{R}}^{\sigma \dagger}a_{\mathbf{R}'}^{\sigma} \nonumber\\ & + \frac{1}{2}\sum_{\sigma \rho}\sum_{\mathbf{R}\mathbf{R}'}U_{\mathbf{R}\mathbf{R}'}a_{\mathbf{R}}^{\sigma \dagger}a_{\mathbf{R}'}^{\rho \dagger}a_{\mathbf{R}'}^{\rho}a_{\mathbf{R}}^{\sigma},
\end{align}
in the case of a single band.  
In this representation of a system, the hopping and Coulomb terms may vary across the lattice. This Hamiltonian represents the system
within a low-energy subspace, typically around 
the Fermi level. Here $\mathbf{R}=\mathbf{R}'$ represents on-site interactions, whereas $\mathbf{R}\neq\mathbf{R}'$ refers to longer-range interactions. While here we will restrict ourselves to including nearest-neighbor terms, one could in principle account for
longer-range interactions as well. 

Solving for the ground and excited states
of the Hubbard model is a quantum many-body
problem, which generally scales
exponentially with the lattice size when solved exactly on classical computers. As a result, one is restricted to studying
finite clusters of relatively small sizes~\cite{dagottoStaticDynamicalProperties1992},
although there exist several approximate methods and numerical techniques which
have allowed studying larger systems~\cite{PhysRevX.5.041041, qinHubbardModelComputational2022,mejuto2019dynamical,PhysRevB.80.075116,RevModPhys.77.1027}. On the other hand, the Hamiltonians of Eq.\,\eqref{eq:basic_Hubbard} and Eq.\,\eqref{eq:Hamiltonian} are amenable to simulation on quantum computers,  where a linear scaling with system size is possible in the ideal scenario~\cite{Wecker15}. As quantum
computers progress into the (partially) fault-tolerant
era, several benefits are expected from
solving increasingly complex versions of the
Hubbard model~\cite{agrawal2024quantifyingfaulttolerantsimulation,dutkiewicz2024errormitigationcircuitdivision}. These facts make quantum computing an attractive avenue to pursue for tackling such problems. Indeed, there have been several studies solving different
flavors of the Hubbard model on quantum computers or classical
simulators thereof~\cite{Cade2020,PhysRevA.93.032303,Reiner2019,caiResourceEstimationQuantum2020, Stanisic2022, anselmemartinSimulatingStronglyInteracting2022, minehSolvingHubbardModel2022,yaoGutzwillerHybridQuantumclassical2021,mukherjeeComparativeStudyAdaptive2023,arute2020observation,PhysRevApplied.23.044028}.

A widely used family of methods for quantum simulation on noisy intermediate-scale quantum (NISQ) hardware~\cite{gustafson2024,sorourifar2024}
are variational quantum algorithms (VQAs)~\cite{Cerezo2021}, where generally, a trial wavefuction
dependent on variational parameters is updated
iteratively in order to minimize a cost function. 
Several studies have shown that
VQAs can reproduce qualitative
features of the Hubbard model
such as its magnetic properties~\cite{Stanisic2022,anselmemartinSimulatingStronglyInteracting2022}, and it
has been shown that for a $1\times 8$ Hubbard chain increasing the ansatz complexity can lead to well-converged values for energies and site occupancies, while spin-spin correlation functions are more challenging
to capture~\cite{anselmemartinSimulatingStronglyInteracting2022}. However, VQAs are also known to
face important challenges, including the 
barren plateau problem, where minimizing 
a cost function can become exponentially difficult
with the depth of a quantum circuit, due to
vanishing gradients~\cite{Wang2021,larocca2024reviewbarrenplateausvariational}. 
At this point, and given the well-known challenges associated with VQAs, it remains unclear whether the qualitative success of these methods in describing features of the Hubbard model will also translate into the quantitative reproduction of ground state properties such as the energy or correlation functions, and whether this will depend on the system size and other factors. Understanding the quantitative characteristics of the VQA solutions of Hubbard models is particularly timely as the
field is transitioning from the NISQ and towards the fault-tolerant era, where systems 
of increasing complexity may be simulated. 
It is therefore important to systematically benchmark the impact of varying lattice sizes on the quantitative features of the solutions obtained with VQAs, particularly given the importance of extrapolations to the thermodynamic limit for drawing comparisons to established methods~\cite{PhysRevX.5.041041}. 
Moreover, it is crucial to establish the impact of electronic correlations, the chosen wavefunction ansatz, and the presence of off-site Coulomb interactions and spatially inhomogeneous parameters on the accuracy of VQAs, as we are moving towards the quantum simulation of extended Hubbard models that
represent real materials. 

Here we present a detailed classical benchmarking study of the variational quantum eigensolver~\cite{Cerezo2022,TILLY20221} (VQE), a promising algorithm for quantum simulation on NISQ hardware~\cite{gustafson2024,sorourifar2024}, applied to single-band Hubbard models of varying degrees of complexity. 
We perform classical tensor-network simulations of VQE calculations for Hubbard models at half filling on 1D and 2D lattices of up to 16 sites, corresponding to 32 qubits. 
We consider different ans\"atze, onsite interaction strengths $U/t = 2, 8$ and the inclusion of nearest-neighbor interactions $V$ and spatial inhomogeneities. 
We show that the so-called number preserving (NP) ansatz, designed specifically for Hubbard models~\cite{Cade2020}, outperforms other popular ans\"{a}tze, but still incurs errors and converges extremely slowly for larger lattices with strong electronic correlations. 
Moreover, we demonstrate that VQE with the NP ansatz can describe Hubbard models with nonuniform parameters and interactions beyond the on-site term with a similar accuracy to the description of uniform models including only on-site terms, which is encouraging for the utility of this approach for solving Hamiltonians representing complex materials~\cite{PhysRevApplied.23.044028}. 
Additionally, we show that an optimization based on maximizing the overlap with a wavefunction computed with a classical reference method can substantially improve the wavefunction fidelity and capture long-range correlations.
Our detailed benchmarking study serves as a reference point along the path to accurate simulation of quantum materials on quantum computers.

\section{Methods}
\label{sec:Methods}
\subsection{Classical simulation of VQE}
Our aim is to obtain the ground state of generalized Hubbard models with Hamiltonians of the form in Eq.\,\eqref{eq:Hamiltonian}. Here, we focus on models with and without nearest-neighbor repulsion and spatially anisotropic hopping parameters. To do so, we simulate the VQE classically by representing the variational ansatz state as a matrix product state (MPS) within a recently proposed variational tensor network eigensolver (VTNE) approach~\cite{Khan2023}. Specifically, following Ref.~\cite{Khan2023}, we start from a product state $\ket{\psi_0}$ in a checkerboard configuration, \emph{i.e.}, with alternating spin up and down. We then generate a variational ansatz state via a parameterized quantum circuit (PQC) as follows:
\begin{equation}
    \label{eq:PQC}
    \ket{\psi_{\mathrm{PQC}}(\bm{\theta})}=U_n(\bm{\theta}_n)...U_1(\bm{\theta}_1)\ket{\psi_0}.
\end{equation}
The precise form of the operators $U_i$ ($i=1,\dots,n$) is determined by the choice of ansatz used in our simulations, more details on which are given below. Each of these operators takes
as arguments a set of parameters $\bm{\theta}_i$, which are initialized randomly. Some ans\"atze also include additional parameter-free gates as specified below. The PQC is represented as an MPS $\ket{\psi_{\chi}(\bm{\theta})}$ with bond dimension $\chi$, and we can therefore compute the energy expectation value
\begin{equation}
    \label{eq:energy}
    E_{\chi}(\bm{\theta})=\bra{\psi_{\chi}(\bm{\theta})}H\ket{\psi_{\chi}(\bm{\theta})},
\end{equation}
with the Hamiltonian $H$ represented as a matrix product operator (MPO). 
Within our optimization scheme, we vary the parameters $\bm{\theta}$ of the PQC in order
to minimize the energy in Eq.\,\eqref{eq:energy}. The gradient of
the energy, which is used to drive the optimization, is computed within the VTNE
scheme as outlined in Ref.~\cite{Khan2023}. Briefly,  starting from the final state $\ket{\psi_{\text{PQC}}(\vec{\theta})}$, the gradient with respect to a parameter $\theta_k$ is computed as
$
\frac{\partial E}{\partial \theta_k} = 2\Re \bra{ \psi_L^{(k)} } \partial_{\theta_k} U_k^\dagger \ket{ \psi_R^{(k)}},
$
where $\ket{\psi_L^{(k)}}$ and $\ket{\psi_R^{(k)}}$ are intermediate states $
\bra{\psi_L^{(k)}} = \bra {0}U_1^\dagger U_2^\dagger\ldots  U_{k-1}^\dagger  $, $
\ket{\psi_R^{(k)}} = U_{k+1}^\dagger \ldots U_n^\dagger H\ket{\psi_{\text{PQC}}(\vec{\theta})}$. The derivative can then
be computed iteratively for each parameter $\theta_k$ by recursively updating the intermediate states as $\bra{\psi_L^{(k-1)}}=\bra{\psi_L^{(k)}}U_{k-1}$ and $\ket{\psi_R^{(k-1)}}=U_k^{\dagger}\ket{\psi_R^{(k)}}$.

Unless otherwise explicitly stated, we use the maximum bond dimension $\chi_{\rm max}=2^{n_q/2}$, where $n_q=2 N_x N_y$ is the number of qubits needed to simulate the Hubbard model on an $N_x\times N_y$ square lattice.
Since this bond dimension is sufficient to exactly represent an arbitrary wavefunction on $n_q$ qubits, the MPS representation yields the energy of
the exact PQC. For all tensor operations in this work we have used the ITensor software package~\cite{itensor,itensor-r0.3}. Our VQE results are benchmarked against those obtained within the density matrix renormalization group (DMRG)~\cite{Verstraete2023} solution of the studied systems, as implemented within ITensor. 

Similar to Ref.~\cite{Khan2023}, we first optimize
the non-interacting ($U=0$) case, and
then use the resulting parameters $\bm{\theta}$ to initialize the
optimization of the full interacting case. The random initial parameters $\bm{\theta}$ are obtained from a Gaussian distribution $\mathcal{N}(0,10^{-5})$ with zero mean and variance $\sigma^2 = 10^{-5}$. The energy minimization is terminated when one of three conditions is satisfied: the energy tolerance~\footnote{The energy tolerance is defined by the absolute difference between the energy at the final step and the energy at the penultimate step.}  reaches $10^{-7}$, the energy gradient reaches $10^{-6}$, or the optimization reaches $1000$ steps. For every value of $U/t$ and for each lattice size, we perform ten independent optimizations (using the L-BFGS method~\cite{nocedal1999numerical}), starting from different random parameters $\bm{\theta}$, and we take the minimum value among these as our estimate of the ground state energy. In this manner, we reduce the chances of the system becoming trapped in a local minimum. 

For some cases studied here, we compare the results from energy-based optimization to those from an overlap-based optimization, which minimizes the following loss function, defined as the logarithm of the infidelity of the variational wavefunction \( \ket{\psi_{\text{PQC}}(\vec{\theta})} \) with respect to the DMRG wavefunction \( \ket{\psi_{\rm DMRG}} \), considered as the ground truth:

\begin{equation}
    \label{eq:loss_overlap}
    f=\log_{10}(1-|\bra{\Psi_{\text{PQC}}(\vec{\theta})}\ket{\Psi_{\rm DMRG}}|^2).
\end{equation} 
To evaluate the gradient of \(f\) with respect to a variational parameter $\theta_k$, we again use the intermediate states $\ket{\psi_L^{(k)}}$ and $\ket{\psi_R^{(k)}}$, which isolate the contribution of the gate parameter $\theta_k$ during the gradient calculation. The gradient contribution from each gate in the ansatz is computed as:
\begin{align}
\frac{\partial f}{\partial \theta_k} = -\frac{2}{IF \ln(10)} \nonumber \\
\times \text{Re}\Bigg( 
\bra{\psi_L^{(k)}} \partial_{\theta_k} U_k^\dagger \ket{\psi_R^{(k)}}
 \bra{\psi_{\rm DMRG}}\ket{\psi_{\text{PQC}}(\vec{\theta})}\Bigg) 
%\nonumber 
%\\- \bra{\psi_L^{(k)}} \partial_{\theta_k} U_k^\dagger \ket{\psi_L^{(k)}}
%|\bra{\psi_{\text{PQC}}(\vec{\theta})}  \ket{\psi_{\rm DMRG}}|^2
%\Bigg),
\end{align}
where $IF=1-|\bra{\Psi_{\text{PQC}}(\vec{\theta})}\ket{\Psi_{\rm DMRG}}|^2$. Once again, the intermediate states are updated iteratively, as discussed for
the gradient calculation of the energy. 

Comparing the performance of the energy- and overlap-based approaches helps to disentangle the effects of cost function landscape and ansatz expressivity in capturing the physics of the Hubbard model.  
Of course the ultimate goal of VQAs is to
obtain the ground state properties of systems for which classical methods would
struggle, which would in turn constitute
such an overlap-based optimization impractical. However, even in those cases
one could perform an overlap-based optimization using an approximate, but still reasonably accurate classically-computed ground state for the system of interest. This would in turn provide a good starting point for
subsequent optimization based on energy minimization, making it important to 
benchmark the performance of this overlap-based approach. 

Before we proceed, it is worth emphasizing that, while the VTNE approach is generally meant as a pre-optimization for VQE~\cite{Khan2023}, our simulations use the maximal bond dimension $\chi_{\rm max}$ (unless otherwise stated, such as for some of our 32 qubit results) and therefore constitute an exact simulation of the VQE algorithm for a given PQC.
We will therefore simply refer to our VTNE optimization as VQE. 

\subsection{Ans\"{a}tze}
\label{sec:ansaetze}
Below we outline the ans\"{a}tze for the PQC in Eq.\,\eqref{eq:PQC} which are used throughout this manuscript.

\subsubsection{Number preserving (NP) ansatz}

The number preserving (NP) ansatz is a generalization of the Hamiltonian Variational Ansatz (HVA)~\cite{Wecker15a}, and it was
designed in Ref.~\cite{Cade2020} specifically within the context of
finding the ground state of the Hubbard model, making it a natural choice for our benchmarking study. 
As discussed in Ref.~\cite{Cade2020}, for a 2D square-lattice system, a layer of the HVA applies a unitary operator of the form
\begin{equation}
    \label{eq:HVA}
    e^{it_{v_2}H_{v_2}}e^{it_{h_2}H_{h_2}}e^{it_{v_1}H_{v_1}}e^{it_{h_1}H_{h_1}}e^{it_{H_o}H_{o}},
\end{equation}
where $t_{H_o},t_{h_1},t_{h_2},t_{v_1},t_{v_2}$ are variational parameters.
Additionally, $H_o$ consists of the on-site terms of the Hamiltonian. 
The vertical hopping terms in the Hamiltonian are partitioned into components $H_{v_1}$,$H_{v_2}$ such that each component consists of mutually commuting terms, and the same is done with the horizontal hopping terms, where the partitions are denoted as $H_{h_1}$,$H_{h_2}$. 
The NP ansatz replaces each hopping and on-site term with a general two-qubit NP operator, which takes two parameters $\theta,\phi$:
\begin{equation}
    \label{eq:NP_matrix}
    U_{\rm NP}(\theta,\phi)=\begin{pmatrix}
1 & 0 & 0 & 0\\
0 & \cos\theta & i\sin\theta & 0\\
0 & i\sin\theta & \cos\theta & 0\\
0 & 0 & 0 & e^{i\phi}\\
\end{pmatrix}.
\end{equation}
This generalized HVA introduces independent parameters for each term in the Hamiltonian, unlike the original HVA~\cite{Wecker15a}, which utilizes a single parameter for each class of terms, \emph{e.g.}, for all horizontal hopping terms. Therefore, while for the NP ansatz used here the number of parameters in the optimization scales with the system size, this is not the case in the original HVA. 
Specifically, a single layer of this ansatz consists of applying the $U_{\text{NP}}$ gate to each pair of qubits coupled by the Hamiltonian of Eq.\,\eqref{eq:basic_Hubbard}, and requires $2[2(N_x(N_y-1)+N_y(N_x-1))+N_xN_y]=10N_xN_y-4N_x-4N_y$ parameters for an $N_x \times N_y$ lattice. Note that the number of parameters in the NP ansatz scales to leading order as $\mathcal{O}(\ell N_x N_y)$, where $\ell$ is the number of layers. In contrast, the number of parameters scales as $\mathcal{O}(\ell)$ in the original HVA of Ref.~\cite{Wecker15} (see Eq.~\eqref{eq:HVA}). Similar to Ref.~\cite{Khan2023}, we apply $R_z(\theta)$ gates
to each qubit prior to the application of the NP
ansatz, which was found to improve optimization. This leads to a total of $(10N_xN_y-4N_x-4N_y)\ell+2N_xN_y$ variational parameters. 

\begin{figure}[tb]
    \centering    \includegraphics[width=\linewidth]{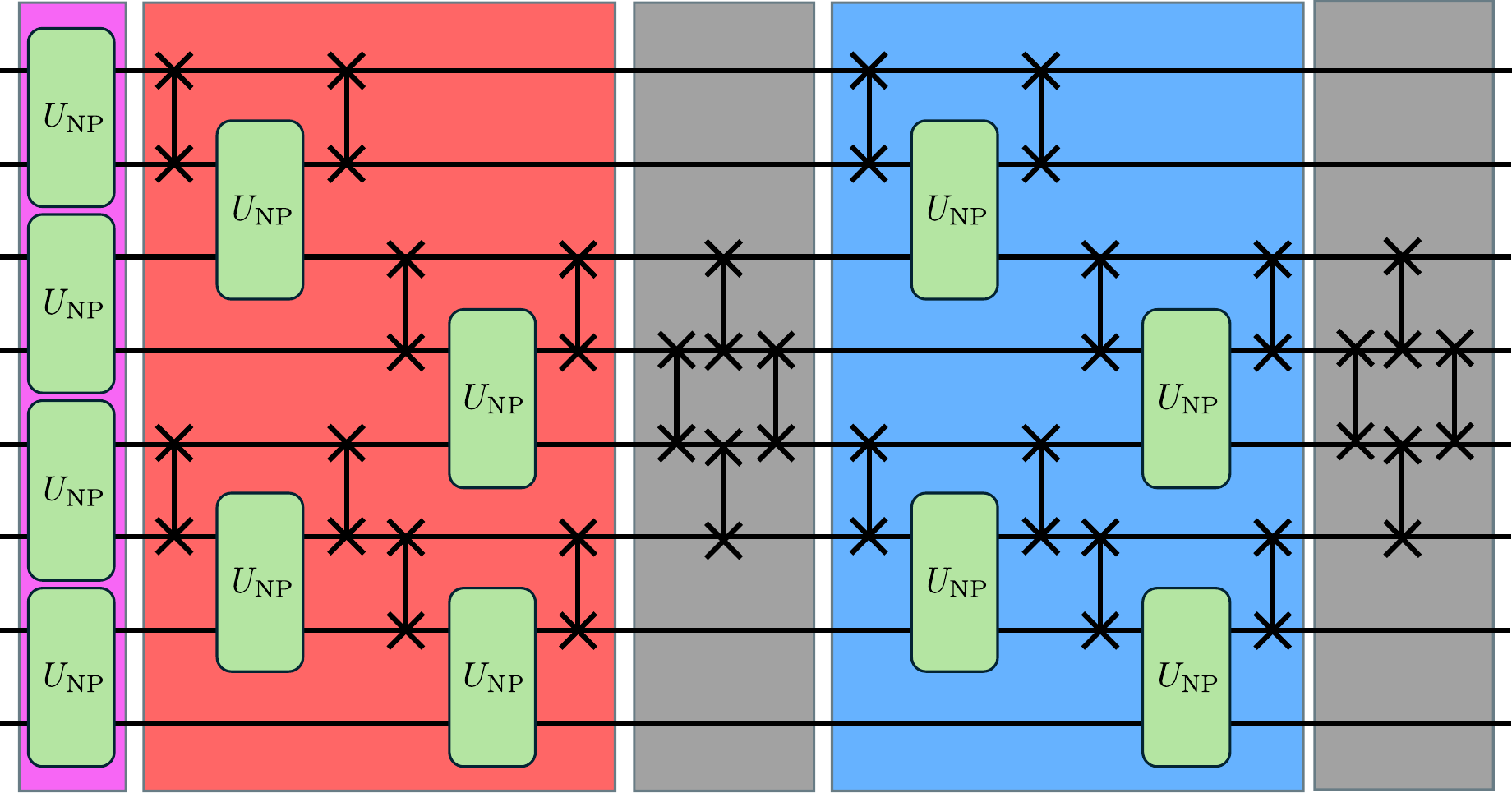}
    \caption{A single layer of the NP ansatz for a $2\times 2$ lattice. Following the application of the on-site terms (purple block), the horizontal hopping terms are implemented (red), followed by fermionic swaps (gray), and subsequently by vertical hopings (blue), and another series of fermionic swaps.}
    \label{fig:NP_ansatz}
\end{figure}

The qubit encoding for the NP ansatz uses a Jordan-Wigner transformation, where each spin-orbital is represented by a qubit. The qubits are indexed using a row-major ordering, where the index of a site $(x,y)$ is given by $(x-1)N_y + y$. Since there are two spin species per site, the qubits are ordered with spin-up and spin-down components stored consecutively for each lattice site. This ordering simplifies the circuit construction but introduces a challenge for implementing hopping terms between physically adjacent sites in the 2D lattice, as they may not be directly adjacent in the qubit chain. 
After we apply the NP operator, we apply fermionic SWAPs to send the qubits back. 
As discussed in Ref.~\cite{Cade2020,hagge2022optimalfermionicswapnetworks}, a simplified circuit has hopping terms between vertically adjacent qubits
implemented locally using an NP gate by applying a parameterless fermionic SWAP gate 
\begin{equation}
    \label{eq:FSWAP}
    \text{FSWAP}=\begin{pmatrix}
1 & 0 & 0 & 0\\
0 & 0 & 1 & 0\\
0 & 1 & 0 & 0\\
0 & 0 & 0 & -1\\
\end{pmatrix}.
\end{equation}
Even in the 1D case, fermionic SWAP gates are necessary to ensure that hopping only occurs between qubits representing the same spin species, preserving the correct fermionic anticommutation relations. The quantum circuit
corresponding to a single layer of the NP
ansatz is schmeatically given in Fig.\,\ref{fig:NP_ansatz},
for the case of a $2\times 2$ lattice.

\subsubsection{Excitation preserving (EP) ansatz}

\begin{figure}[tb]
    \centering    \includegraphics[width=0.8\linewidth]{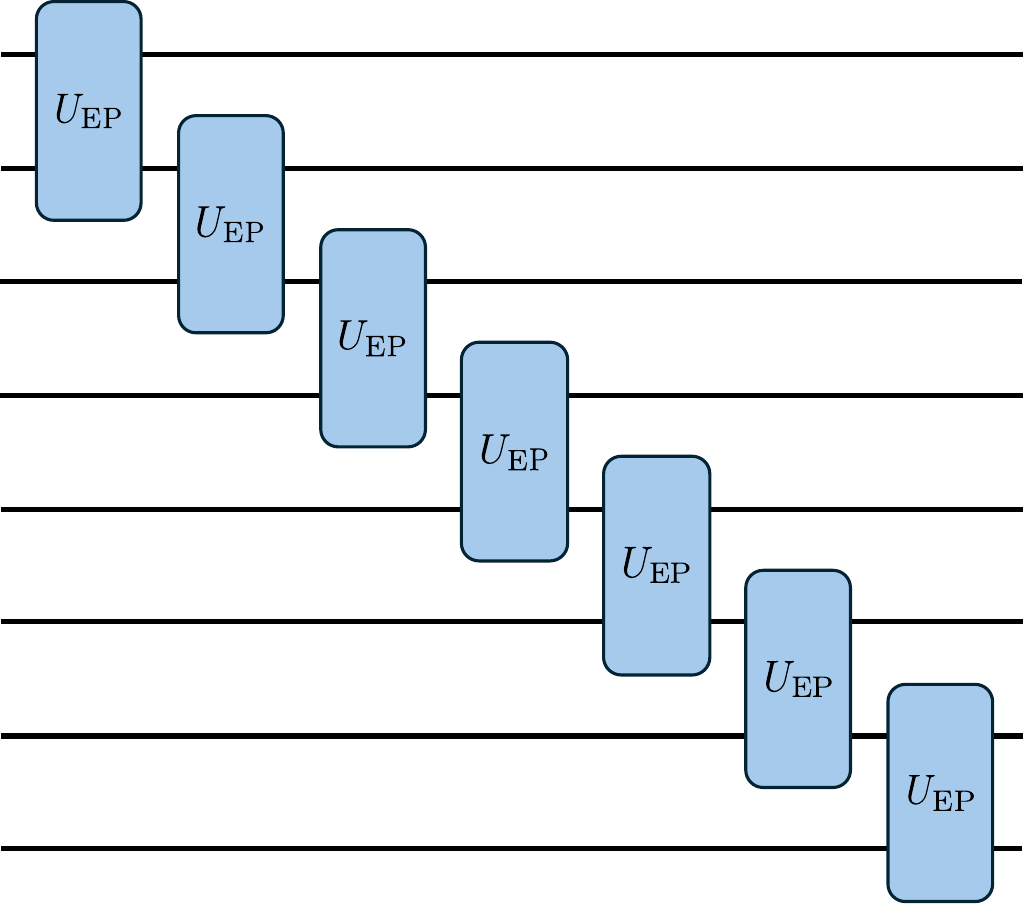}
    \caption{A single layer of the EP ansatz for a $2\times 2$ lattice.}
    \label{fig:ΕP_ansatz}
\end{figure}

We also employ an excitation preserving (EP) ansatz,
which has certain similarities to the NP ansatz outlined above. As implemented within Qiskit~\cite{Qiskit}, we apply the two-qubit gate
\begin{equation}
    \label{eq:EP_matrix}
    U_{\rm EP}(\theta,\phi)=\begin{pmatrix}
1 & 0 & 0 & 0\\
0 & \cos(\theta/2) & -i\sin(\theta/2) & 0\\
0 & -i\sin(\theta/2) & \cos(\theta/2) & 0\\
0 & 0 & 0 & e^{-i\phi}\\
\end{pmatrix}.
\end{equation}
This is very similar to the NP operator of Eq.\,\eqref{eq:NP_matrix}. 
However, unlike the NP case, we simply apply the gate to neighboring
qubits according  to the indexing scheme where the index of a site $(x,y)$ is given by $(x-1)N_y + y$, without each term necessarily reflecting the physical hopping and on-site terms of the Hubbard Hamiltonian. This makes the EP ansatz more
generic and straightforward to use compared to NP. However, the EP ansatz is also not tailored
specifically to the physics of the Hubbard
model on a lattice, as will also be reflected by our results. 
Within the EP ansatz, no fermionic swaps are performed, as is done
within the NP ansatz in an effort to replicate
vertical and horizontal hopping terms of
the Hubbard Hamiltonian, between same spin
qubits. Before and after the application of $U_{EP}$ we apply single-parameter $R_z$ gates, leading
to a total number of $4N_xN_y+2(2N_xN_y-1)\ell$ variational parameters for this generic ansatz. The quantum circuit representing a single layer of the EP ansatz for a $2\times 2$ lattice is visualized in Fig.\,\ref{fig:ΕP_ansatz}.

\subsubsection{Unitary coupled cluster (UCC) ansatz}
The unitary coupled cluster (UCC) ansatz 
consists of
applying excitation operators to the initial
wavefunction $\ket{\psi_0}$ as follows:
%is the exponentiation of the coupled cluster operator $\hat{T}$ acting on the Hartree--Fock reference wave function $\ket{\Psi_0}$,
\begin{gather}
    \ket{\Psi_{\mathrm{UCC}}} = \exp( \hat{T} - \hat{T}^{\dagger})\ket{\Psi_{0}} , \\
    \hat{T} = \sum_{i}^{\mathrm{occ}} \sum_{a}^{\mathrm{vir}} \theta_{i}^{a}\hat{a}_{a}^{\dagger}\hat{a}_{i}
            + \sum_{ij}^{\mathrm{occ}} \sum_{ab}^{\mathrm{vir}} \theta_{ij}^{ab}\hat{a}_{a}^{\dagger} \hat{a}_{b}^{\dagger} \hat{a}_{j} \hat{a}_{i} + \cdots \ .
\end{gather}
Here the creation and annihilation operators $\hat{a}^{\dagger}$ and $\hat{a}$ act on the occupied ($i,j,\ldots$) and virtual ( $a,b,\ldots$) orbitals of the initial wavefunction, respectively.

We employ the factorized form of the UCC ansatz,
\begin{equation}
    \label{eq:ansatz}
    \ket{\Psi_{\mathrm{UCC}}} = \prod_{ij\cdots}^{\mathrm{occ}} \prod_{ab\cdots}^{\mathrm{vir}} \hat{U}^{ab\cdots}_{ij\cdots}\ket{\Psi_{0}},
\end{equation}
for which the individual gates are defined as
\begin{equation}
    \label{eq:factor}
    \begin{gathered}
    \hat{U}^{ab\cdots}_{ij\cdots} = \exp(\theta_{ij\cdots}^{ab\cdots}(\hat{a}_{ij\cdots}^{ab\cdots} - \hat{a}_{ab\cdots}^{ij\cdots})) \\
    \hat{a}_{ij\cdots}^{ab\cdots} = \hat{a}^\dagger_a\hat{a}^\dagger_b \dots \hat{a}^{\phantom{\dagger}}_j\hat{a}^{\phantom{\dagger}}_i.
    \end{gathered}
\end{equation}
We only include single- and double-excitation operators, $\hat{U}_i^a$ and $\hat{U}_{ij}^{ab}$, within the so-called unitary coupled cluster ansatz with singles and doubles (UCCSD). The UCCSD ansatz, which we will refer to
simply as the UCC ansatz, is number preserving, similar to the NP and EP ans\"{a}tze, however, unlike the linear scaling of the number of parameters of those ans\"{a}tze
with system size (per layer), UCCSD scales quadratically~\cite{mcclean2016theory}, 
becoming more complex for larger systems.

\section{Results}

\begin{figure*}[tb]
    \centering
    \includegraphics[width=0.8\linewidth]{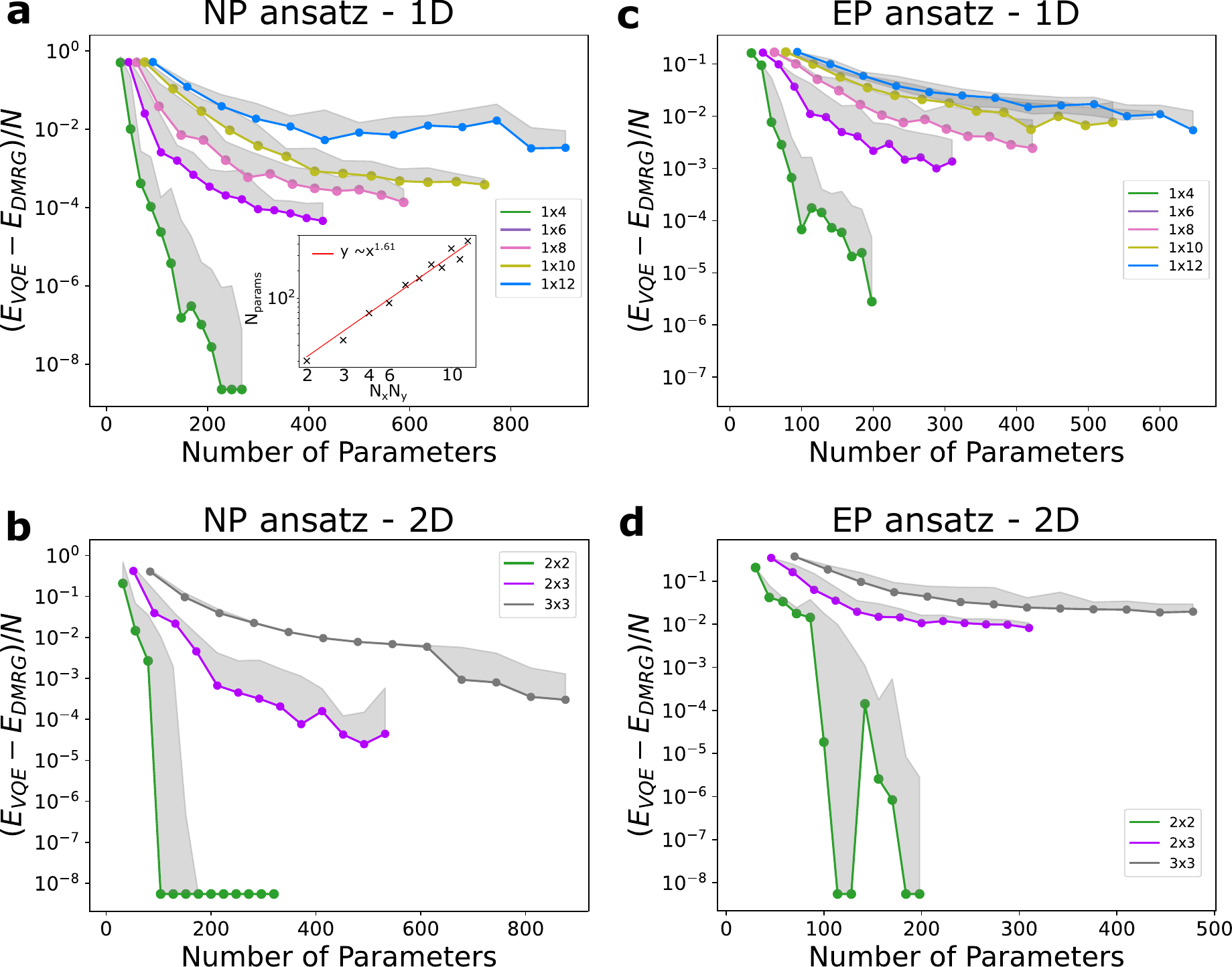}
    \caption{Energy difference per lattice site, between VQE and DMRG ground state energies when using the NP and EP ans\"{a}tze, for 1D and 2D lattices, on a logarithmic scale. The shaded region indicates the difference between the minimum and maximum ground state energies obtained using VQE with a total of ten different random starting points. The inset of panel \textbf{a} shows a log-log plot of the minimum number of parameters needed to achieve an error per lattice site $\Delta=0.01$, when using the NP ansatz for 1D lattices.}
    \label{fig:ansatz}
\end{figure*}

We now systematically study the impact of varying the number of variational parameters and the ansatz, the strength 
of the electronic interactions, and the
strength of disorder and off-site Coulomb interactions, on the quality of the VQE solutions of the half-filled Hubbard model with respect to DMRG. Moreover,
we benchmark the performance of using
an overlap-based optimization, compared to
the more conventional energy-based optimization. The optimized VQE energies
for all cases presented below are given
in the Supplemetary Material.

\subsection{Impact of the ansatz and the number of variational parameters}

We start our analysis by describing the VQE result for the half-filled Hubbard model of different system sizes in 1D and 2D square lattice geometries. We employ open boundary conditions throughout. We will first explore
how the NP ansatz, which was designed specifically for solving the Hubbard model on quantum hardware, performs as a function of lattice size and number of ansatz parameters. 
Our benchmarks make use of PQCs of up to thirteen layers. 

In Fig.~\ref{fig:ansatz}a we plot for several one-dimensional lattices the difference between our VQE energy and the DMRG benchmark per lattice site, for $U/t=2$.
This interaction strength lies in the weakly- to moderately-correlated regime~\cite{qinBenchmarkStudyTwodimensional2016, qinHubbardModelComputational2022}. We note that in the thermodynamic limit both 1D and 2D Hubbard models for $U>0$ are in a Mott insulating phase at half filling with dominant antiferromagnetic N\'eel-type correlations~\cite{PhysRevB.80.075116,giamarchi2003quantum}. While the charge gap is nonzero, the spin gap vanishes. As our ground state energy from VQE we plot our best estimate, obtained as the lowest value from 
ten independent optimizations with random starting points. The shaded regions in Fig.~\ref{fig:ansatz}a indicate the full range of energies obtained over all optimizations. 
In Fig.~\ref{fig:ansatz}a we
only visualize the results for even
numbers of sites, however, in the Supplemental Material we also provide
the results for odd lattices, which are consistent with the trends observed here. 

While it is clear that the VQE optimization converges to the ground state energy even for the larger lattices as more parameters are added to the PQC, the convergence becomes increasingly slow for larger lattice sizes, and even plateaus to errors of order $10^{-2}$ in those cases. For the $1\times 12$ lattice
with $13$ layers of the NP ansatz we have
also performed an extended optimization, where we performed a hundred independent optimizations (rather than ten) starting from different random parameters. Nevertheless, even when we explore a larger part of phase space, we do not find any appreciable reduction in the ground state energy produced by VQE.
Fig.~\ref{fig:ansatz}b visualizes the
performance of VQE on a logarithmic scale for two-dimensional lattices at $U/t=2$.
We again see that the convergence of VQE using the NP ansatz slows down with increasing system size, consistent with our benchmarks of the one-dimensional case. 
This plateauing is the result of the complex optimization landscape of larger systems with a significant number of variational parameters, which makes VQE prone to becoming stuck in local minima. This suggests further work could be devoted to using a combination of optimizers including genetic algorithms.
While the energy gradients during the course of our optimizations generally always remain finite and we do not encounter barren plateaus, these could
appear and pose a challenge on quantum hardware, where the numerical accuracy of
gradients is limited by shot noise.

To further understand the convergence behavior of the
VQE energy towards the DMRG value, we plot in the inset of Fig.~\ref{fig:ansatz}a 
the minimum number of parameters that is required in
order to reach an error value $\Delta=\frac{E_{\text{VQE}}-E_{\text{DMRG}}}{N}$ of
$0.01$ with the NP ansatz and for 1D lattices. We see that the necessary number of parameters
to reach $\Delta=0.01$ scales as $(N_x \times N_y)^n$,
with $n=1.61$. A similar power-law behavior with an exponent $n$ of $2.13$ is found for 2D
systems, when we aim to reach the same
error value. It should be emphasized that
this power-law behavior is established empirically through our analysis of the numerical results we obtain in this work,
nevertheless, it will be useful for anticipating the computational cost of VQE simulations for larger systems. 

\begin{figure}[tb]
    \centering    \includegraphics[width=0.8\linewidth]{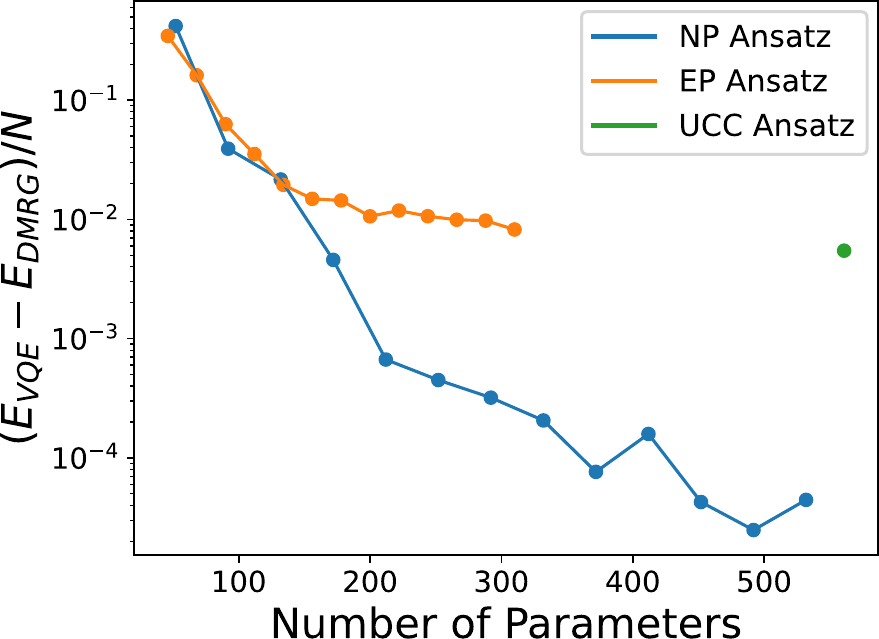}
    \caption{Energy difference per lattice site, of the VQE ground state from DMRG for a $2\times 3$ lattice, as a function of the number of parameters in the NP and EP ans\"{a}tze (corresponding to 1 to 13 ansatz layers), and also when using the UCC ansatz.}
    \label{fig:NP_vs_EP_vs_UCC}
\end{figure}

We now turn to results obtained with the EP ansatz. 
Figs.~\ref{fig:ansatz}c and \ref{fig:ansatz}d visualize the error per lattice site of VQE with respect to DMRG for $U/t=2$ and for one- and two-dimensional lattices, respectively. Once again, for visualization purposes we only show the results for even one-dimensional lattices, and include all data for odd lattices in the Supplemental Material.
As for the case of the NP ansatz, we optimize PQCs of up to thirteen layers. In all cases, the performance of VQE based on the EP
ansatz is worse than that of VQE based on the NP ansatz. As already discussed, this is unsurprising given the generic character of the EP ansatz, compared to
the NP ansatz where different ansatz components represent specific physical terms of the Hamiltonian. The difference between the two becomes particularly pronounced for two-dimensional lattices, where it is immediately evident that the VQE energy converges to the DMRG one at a much slower rate than when using the NP ansatz.
Moreover, for larger lattices we observe that the VQE energy plateaus at substantial error values; when using $13$ layers of this ansatz for a $3\times 3$ lattice for example, the error per lattice site of the VQE energy compared to the DMRG value is $0.019$-$0.029$, compared to $0.0003$-$0.001$ for the NP ansatz. 
We attribute this behavior to the fact that the EP ansatz does not directly couple vertically adjacent qubits in two-dimensional lattices, due to the lack of fermionic swaps, hence missing additional key parts of the physics of the Hubbard model. We conclude that the performance of EP is  limited by its expressivity. 
Another interesting observation when comparing the layered ans\"atze is that, for small parameter numbers $\lesssim150$, the EP ansatz performs similar to NP. 
However, the rate of convergence with the number of parameters slows down markedly for the EP ansatz as this parameter number threshold is surpassed, and
the NP ansatz systematically yields smaller errors. We find
that it is a general feature
across different lattices that
for smaller number of parameters the NP and EP ans\"atze perform similarly, 
however the energy error obtained from optimizing the EP PQC plateaus for larger numbers of parameters. 

The UCC ansatz is a popular choice for VQE calculations, particularly for chemistry applications. 
Given this popularity, we benchmark in
Fig.~\ref{fig:NP_vs_EP_vs_UCC} the performance of the NP, EP and UCC ans\"{a}tze for the solution of a $2\times 3$
Hubbard model with $U/t=2$ (here we only show the lowest VQE energy obtained for each number of parameters and for each ansatz). 
Unlike the layered NP and EP ans\"{a}tze, the UCC ansatz contains a set number of one- and two-electron excitation operators, and a number of parameters that is determined only by the lattice size. 
We see in Fig.~\ref{fig:NP_vs_EP_vs_UCC} that for our case study, the UCC ansatz requires significantly more parameters (quadratic scaling, see Section~\ref{sec:ansaetze}) compared to NP or EP PQCs of up to thirteen layers (linear scaling),
thus leading to a higher computational cost. 
Indeed we were not able to study lattices larger than $2\times 3$ using the UCC ansatz due to the steep increase in the number of optimization parameters - for a $3\times 3$ lattice, the ansatz has $3213$ parameters, more than three times that of the NP ansatz at thirteen layers. 
Moreover, we see in Fig.~\ref{fig:NP_vs_EP_vs_UCC} that despite the larger number of optimization parameters associated with the UCC ansatz, there is no significant improvement of the results compared to the EP ansatz, while the NP ansatz significantly outperforms it. 
In the future, it will be worthwhile to explore ways to optimize the UCC ansatz for lattice model simulations such as by adaptive approaches or by truncating excitation operators through a ranking scheme of their relative importance, which has been implemented for molecular systems through an MP2 pre-optimization scheme~\cite{Mullinax2023}. 
%Another interesting observation when comparing the layered ans\"atze is that, for small parameter numbers $\lesssim150$, the EP ansatz performs similar to NP. 
%However, the rate of convergence with the number of parameters slows down markedly for the EP ansatz as this parameter number threshold is surpassed, and
%the NP ansatz systematically yields smaller %errors. We find
%that it is a general feature
%across different lattices that
%for smaller number of parameters the NP and EP ans\"atze perform similarly, 
%however the energy error obtained from optimizing the EP PQC plateaus for larger numbers of parameters. 

\begin{figure}[tb]
    \centering \includegraphics[width=\linewidth]{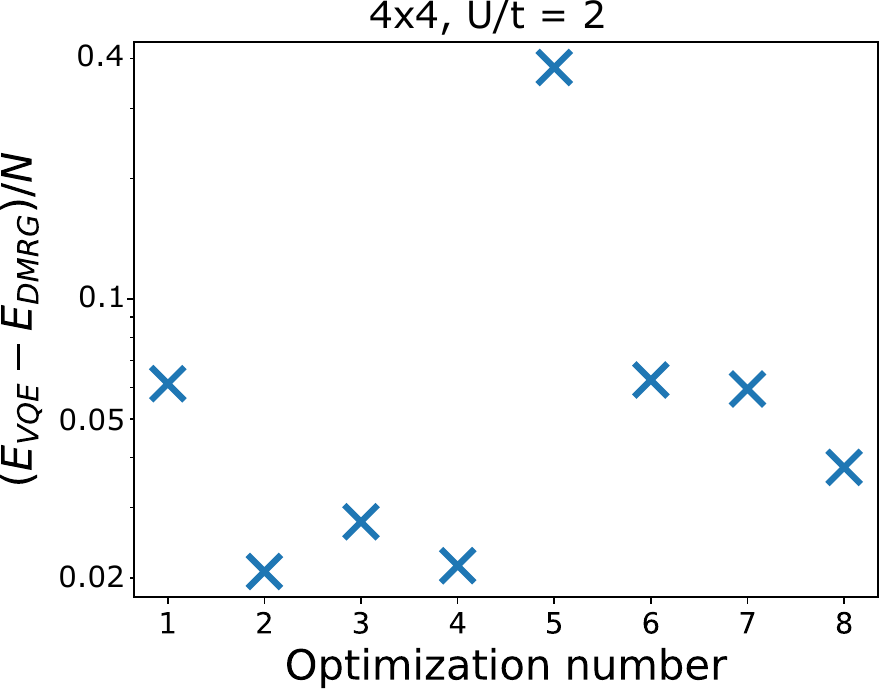}
    \caption{Converged energy difference of VQE and DMRG energy per lattice site, for a $4\times 4$ Hubbard model with $U/t=2$, and for different starting points of the optimization. Both levels of theory employ a bond dimension of $\chi=512$. Here we use $12$ layers of the NP ansatz, corresponding to 1,568 optimization parameters within VQE. }
    \label{fig:NP_2_4x4}
\end{figure}

Our results so far have demonstrated that convergence of the VQE energy to the DMRG value with respect to the number of the optimization parameters becomes slower for larger lattice sizes. 
It is therefore interesting to explore the convergence of the ground state energy in more detail for a larger lattice of size $4\times 4$ for $U/t=2$, using the NP ansatz.
Since each lattice site can be occupied by a spin-up and/or a spin-down electron, this corresponds to $4\times 4\times 2=32$ qubits. Due to the large computational cost associated with simulating a lattice of this size, we restrict the bond dimension of the MPS representation within our VTNE approach, and also within DMRG, to $\chi=512$. In the Supplemental Material we demonstrate the convergence of the DMRG energy per lattice site with the bond dimension, which allows us to estimate that using $\chi=512$ leads to an error of approximately $5\times 10^{-3}$ for this quantity. 
As in previous cases, we initialize ten VQE optimizations from different, random starting points. 
Two of these failed to converge, and for the remaining eight we visualize their converged energy errors per lattice site with respect to DMRG in Fig.~\ref{fig:NP_2_4x4}. 
We see that
the energy errors have a wide distribution, indicative of the fact that some optimizations become stuck
in local minima. Nevertheless, in some
cases we can approach the ground state
energy with reasonable accuracy. To reach this level of accuracy, we allowed each optimization to run for $48$ hours on a single CPU of NERSC Perlmutter, and each utilized
an average of $7.1$ GBs of memory. 
The lowest energy obtained through VQE optimization, with 12 NP ansatz layers corresponding to 1,568 parameters, has a relative error of $2.2\%$ in its energy, compared to the DMRG value. 
This demonstrates that even for these larger lattices, VQE using the NP ansatz can, at least in
principle, yield ground states of
the Hubbard model that are not far
from the ground truth, as long as enough parameters are included in the optimization. We conclude that the NP ansatz is sufficiently expressive, but the optimization is challenging due to local minima. 

\begin{figure}[h!]
    \centering
    \includegraphics[width=0.9\linewidth]{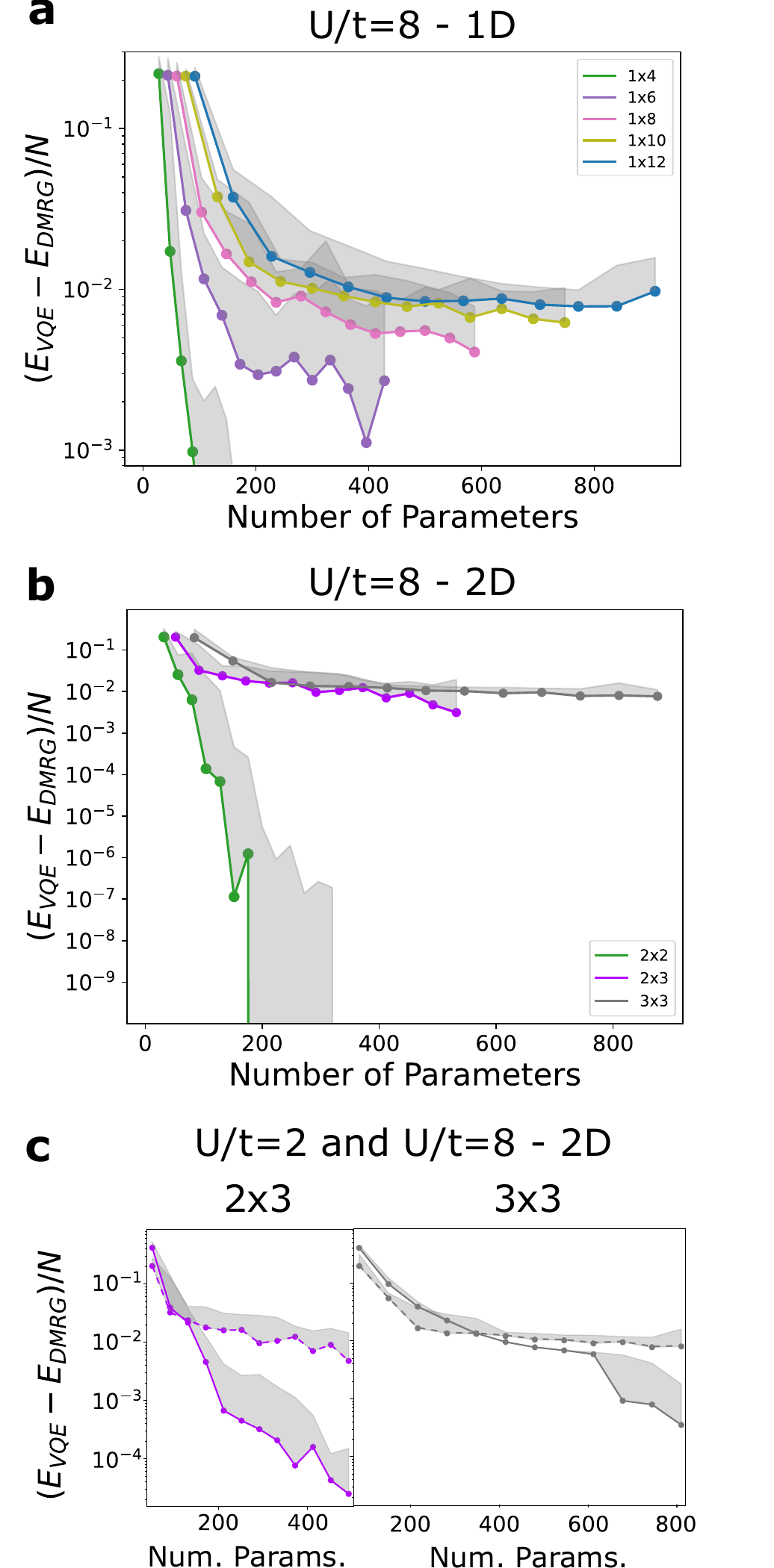}
    \caption{Energy difference per lattice site, of the VQE ground state from DMRG 1D (panel \textbf{a}) and 2D (panel \textbf{b}) lattices, when using the NP ansatz and for $U/t=8$. In panel \textbf{c} we compare the energy error per lattice site of the VQE ground state between the cases of $U/t=2$ (solid lines) and $U/t=8$ (dashed lines) for 2D lattices.}
    \label{fig:interaction}
\end{figure}

\subsection{Impact of interaction strength $U/t$}

\begin{figure*}[tb]
    \centering
    \includegraphics[width=0.8\linewidth]{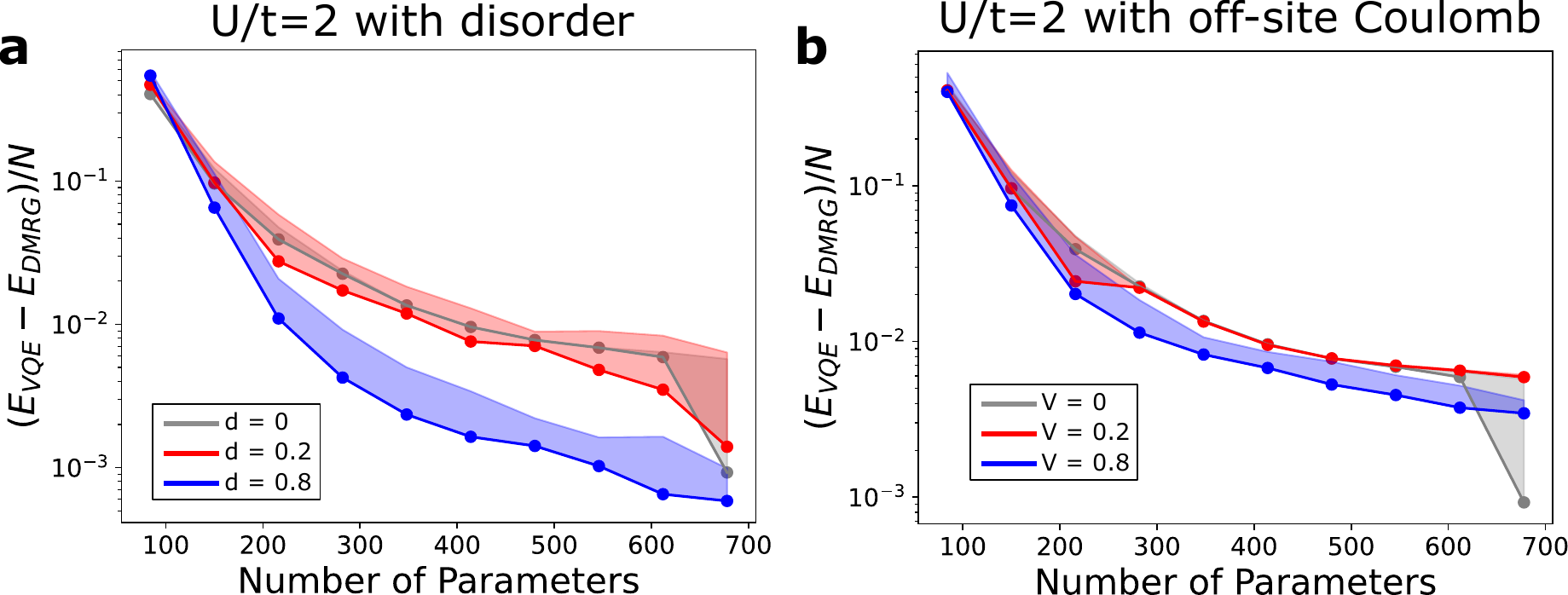}
    \caption{Comparison of the energy difference per site of a $3\times 3$ lattice, of the VQE ground state from DMRG, when using the NP ansatz and for $U/t=2$, in the presence of disorder $d$ (panel \textbf{a}) and %off-site 
    nearest-neighbor (screened) Coulomb interactions $V$ (panel \textbf{b}). The shading represents the range of energies obtained in the different cases.}
    \label{fig:disorder_V}
\end{figure*}

Up to this point, we have
focused on the weakly-correlated case with interaction strength $U/t=2$.
However, for several applications of interest in condensed matter physics, stronger electronic correlations become important, and previous results on $1\times 8$
Hubbard chains indicate that this regime might be more challenging to capture with VQAs~\cite{anselmemartinSimulatingStronglyInteracting2022}.
To address the quality of VQE simulations in this strong coupling regime, we set $U/t=8$. Motivated by the results of Fig.\,\ref{fig:ansatz}, we adopt the NP ansatz in the following.
Figs.~\ref{fig:interaction}a and \ref{fig:interaction}b demonstrate the convergence of the VQE results towards the DMRG ones as a function of the number of parameters in the PQC for various lattice sizes in 1D and 2D, respectively.
It becomes clear from these results and by comparing them to those of Fig.\,\ref{fig:ansatz} that stronger electronic correlations pose more of a challenge to accurate VQE simulations of the Hubbard model, even when using the 
NP ansatz. To emphasize this, we visualize in Fig.\,\ref{fig:interaction}c the VQE energy error per site for $2\times 3$ and $3\times 3$ lattices, for $U/t=2$ (solid lines) and $U/t=8$ (dashed lines) on the same plot.
It is also worth noting that similar to $U/t=2$ and Fig.\,\ref{fig:ansatz}b, we
find for $U/t=8$ that there is a power-law dependence
of the minimum number of parameters needed to achieve an accuracy of $\Delta=0.01$, on the lattice size, with an exponent of $n=2.18$ for 2D systems and $n=1.61$ for 1D systems. 

The challenge posed by systems with greater electronic correlation motivates a search for an ansatz that is tailored to the strong coupling regime, for example, one that exploits the well-known mapping of the Hubbard to the Heisenberg model at strong coupling. Another possibility 
to improve the ground state energies for strongly correlated Hubbard
models could be to use solutions to the Heisenberg model as starting points for VQE, as was recently proposed~\cite{PhysRevB.109.035128}.

\subsection{Impact of disorder and off-site Coulomb interactions}
\label{sec:disorder_V}

An important factor towards increasingly realistic materials simulations on quantum hardware is spatial inhomogeneity not present in the simple Hubbard model of Eq.~\eqref{eq:basic_Hubbard}.
Indeed, the material-specific Hubbard models of the form of Eq.~\eqref{eq:Hamiltonian}, derived through methodologies such as \emph{ab initio} downfolding~\cite{Nakamura2021,Zheng2018,PhysRevB.92.054515,PhysRevApplied.23.044028} and embedding ~\cite{Vorwerk2022,PhysRevX.11.021006}, allow for spatial variations in the hopping and interaction parameters, \emph{e.g.}, due to the presence of different chemical elements or of static disorder within the material. 

We therefore now consider the impact of such disorder on the accuracy of VQE simulations of Hubbard models. 
To do so, we study the ground state properties of the Hamiltonian
\begin{align}
    \label{eq:basic_Hubbard_disorder}H= -\sum_{\sigma}\sum_{\langle \mathbf{R}\mathbf{R}'\rangle}t_{\mathbf{R}\mathbf{R}'}a_{\mathbf{R}}^{\sigma \dagger}a_{\mathbf{R}'}^{\sigma}+ U\sum_{ \mathbf{R}}n_{\mathbf{R}\uparrow}n_{\mathbf{R}\downarrow},
\end{align}
where we now have disorder in the nearest-neighbor hopping ($\mathbf{R} \neq \mathbf{R}'$) and on-site potential ($\mathbf{R} = \mathbf{R}'$) terms, $t_{\mathbf{R}\mathbf{R}'}=t\cdot \delta_{\langle\mathbf{R}\mathbf{R}'\rangle}+\delta t_{\mathbf{R}\mathbf{R}'}$. The function $\delta_{\langle\mathbf{R}\mathbf{R}'\rangle}$ is equal to one for nearest neighbors, and zero otherwise, since the on-site potential in our system is set to zero. 
The disorder enters via $\delta t_{\mathbf{R}\mathbf{R}'}=d\cdot \mathcal{N}(0,1)$,
with $d$ a user-defined parameter for the 
magnitude of the disorder and $\mathcal{N}(0,1)$ a function drawing a random number from a Gaussian distribution of zero mean and unit variance.
The values of $t_{\mathbf{R}\mathbf{R}'}$ for nearest neighbor hopping are centered around $t$, and the on-site potential is centered around zero, both with a standard deviation of $d$. 
In Fig.~\ref{fig:disorder_V}a we plot the error per lattice site of the VQE ground state energies with respect to the DMRG ones, for a $3\times 3$ lattice, and as a function of
the number of optimization parameters in the ansatz. 
We use disorder values of $d=0.2$ (red) and $d=0.8$ (blue).
The results suggest that even for disordered systems with nonuniform parameters, VQE using the NP ansatz can recover the ground state of the Hubbard model with similar accuracy to the disorder-free case $d=0$ (gray). We make similar
findings irrespective of the lattice size, and
we do not include results beyond the $3\times 3$
case in Fig.\,\ref{fig:disorder_V} for visibility purposes. All data for disordered simulations on different lattices are given in the Supplementary Material. Interestingly, we observe in Fig.\,\ref{fig:disorder_V} that the introduction of disorder, particularly for $d=0.8$, actually reduces the
error of the VQE energy with respect
to the DMRG value. One reason behind this could be that the symmetry-breaking effect of disorder lifts degeneracies, effectively simplifying the VQE optimization landscape. Moreover, disorder induces localization, which can reduce entanglement~\cite{PhysRevA.76.042333}, hence making it easier
to capture the ground state with shallower circuits. 

So far we have focused on Hubbard models where Coulomb interactions are only present for electrons on the same site. However, in realistic material systems there are also
longer-range interactions present, which
motivates us to
investigate the impact of including a finite nearest-neighbor repulsive (screened) Coulomb interaction $V$ in our model, \emph{i.e.}, to consider extended Hubbard models of the form
\begin{align}
    \label{eq:basic_Hubbard_V}H= -t\sum_{\sigma}\sum_{\langle \mathbf{R}\mathbf{R}'\rangle}a_{\mathbf{R}}^{\sigma \dagger}a_{\mathbf{R}'}^{\sigma}+ U\sum_{ \mathbf{R}}n_{\mathbf{R}\uparrow}n_{\mathbf{R}\downarrow}\nonumber \\+V\sum_{\sigma \sigma'}\sum_{\langle \mathbf{R}\mathbf{R}'\rangle}n_{\mathbf{R}\sigma}n_{\mathbf{R}'\sigma'}.
\end{align}
In Fig.~\ref{fig:disorder_V}b we visualize
the energy error per lattice site of the VQE solutions
of the extended Hubbard model on a $3\times 3$ lattice, with
respect to the DMRG energy, as a function
of the number of parameters used in the optimization, and for different values of $V$, specifically $V=0$ (gray), $V=0.2$ (red), and $V=0.8$ (blue). Similar to the case of Fig.~\ref{fig:disorder_V}a of having finite
disorder in the lattice, we see that 
having %off-site 
nearest-neighbor interactions
generally leads to comparable errors
in the VQE energy to when $V=0$. 
We make similar findings for all lattice sizes
studied here, with all numerical results given in the Supplementary Material. 
The fact that VQE with the NP ansatz seems capable of describing generalized, extended Hubbard models that include spatial inhomogeneity and (screened) Coulomb interactions beyond on-site, with a similar degree of accuracy as for the 
standard Hubbard model of Eq.\,\eqref{eq:basic_Hubbard}, is encouraging 
for the simulation of increasingly complex, realistic and inhomogeneous systems on quantum hardware. We relate this capability to the fact that, unlike in the original HVA ansatz~\cite{Wecker15}, the NP ansatz contains independent variational parameters on every
bond that appears in the Hamiltonian. 

\begin{figure*}[tb]
    \centering
    \includegraphics[width=\linewidth]{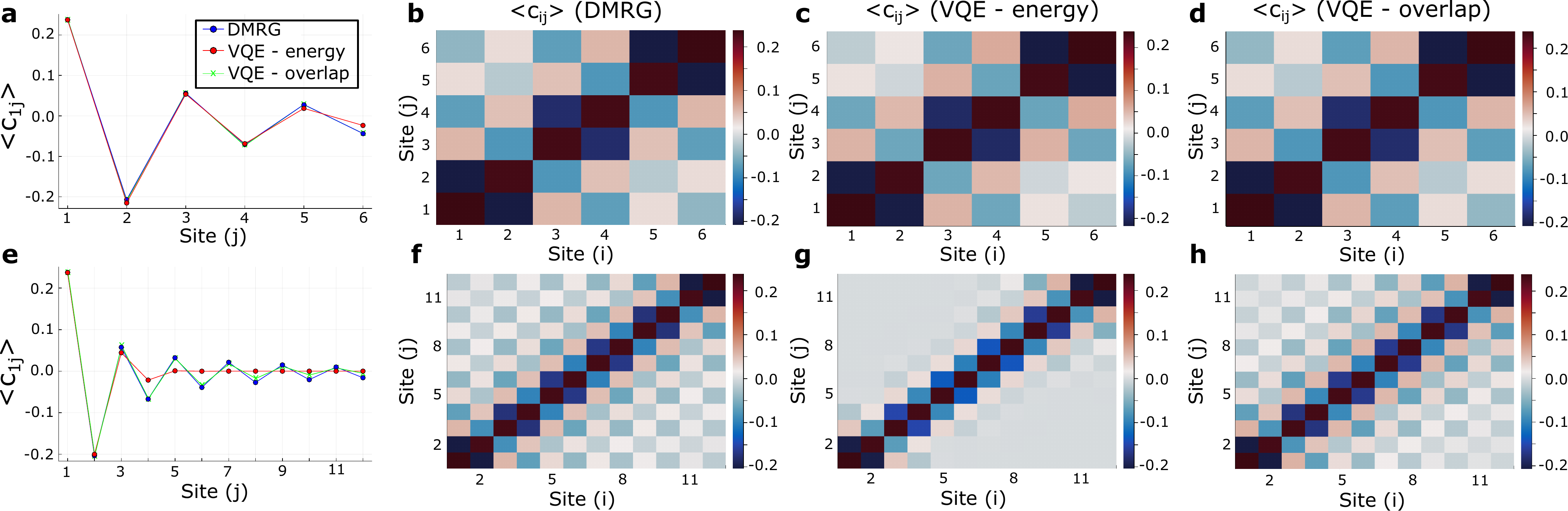}
    \caption{Spin correlation function $\langle C_{1j} \rangle$ (panel \textbf{a}) and  $\langle C_{ij} \rangle$ (panel $\textbf{b}$ at the DMRG level, panel $\textbf{c}$ within energy-based VQE, and panel $\textbf{d}$ within overlap-based VQE), for a $1\times 6$ Hubbard model with $U/t=8$. Panels $\textbf{e}$-$\textbf{h}$ visualize the same quantities for a $1\times 12$ Hubbard chain instead.}
    \label{fig:magnetism}
\end{figure*}

\subsection{Extracting correlation functions and overlap-based optimization}
\label{sec:overlap}
 
Our work so far has focused on the quantitative benchmarking of VQE simulations of the Hubbard model in terms of predicting the ground state energy.
However, qualitative features of the model can be important to reproduce, and indeed VQAs have been shown in various examples to yield ground states that display the well-known antiferromagnetic behavior of the Hubbard model, which is most pronounced in the strong coupling regime~\cite{anselmemartinSimulatingStronglyInteracting2022,PhysRevX.5.041041}. 
To understand the performance of VQE in
terms of recovering such behaviors, we compute the spin correlation function
\begin{equation}
    \langle C_{ij} \rangle = \langle S_i^z S_j^z \rangle - \langle S_i^z \rangle \langle S_j^z \rangle,
\end{equation}
where the braket $\langle \cdot \rangle$ indicates an expectation value, and $S_i^z$ a spin-$z$ operator on site $i$. We visualize the spin correlation function in Fig.~\ref{fig:magnetism}
in the case of $U/t=8$ for a $1\times 6$ and $1\times 12$
chain. We show results for  
VQE based on minimizing the energy,
which has been our focus up to this point,  
and also include results obtained for VQE utilizing overlap-based optimizations as discussed in Section\,\ref{sec:Methods}. We compare the VQE results with DMRG. To obtain the VQE results
here we have used 13 layers of the NP ansatz. 
We observe that for the smaller chain, the spin properties from both VQE approaches and DMRG
are in near perfect agreement.
For the larger $1\times 12$ chain, VQE based on an energy minimization fails to fully capture longer-range spin correlations, in agreement
with previous findings~\cite{Stanisic2022,anselmemartinSimulatingStronglyInteracting2022}. However, the VQE ground state obtained
through overlap-based optimization is
in near-perfect agreement to the DMRG one
in terms of its long-range spin correlation, highlighting the utility 
of this approach for computing such
features. 

\begin{figure*}[tb]
    \centering
    \includegraphics[width=0.8\linewidth]{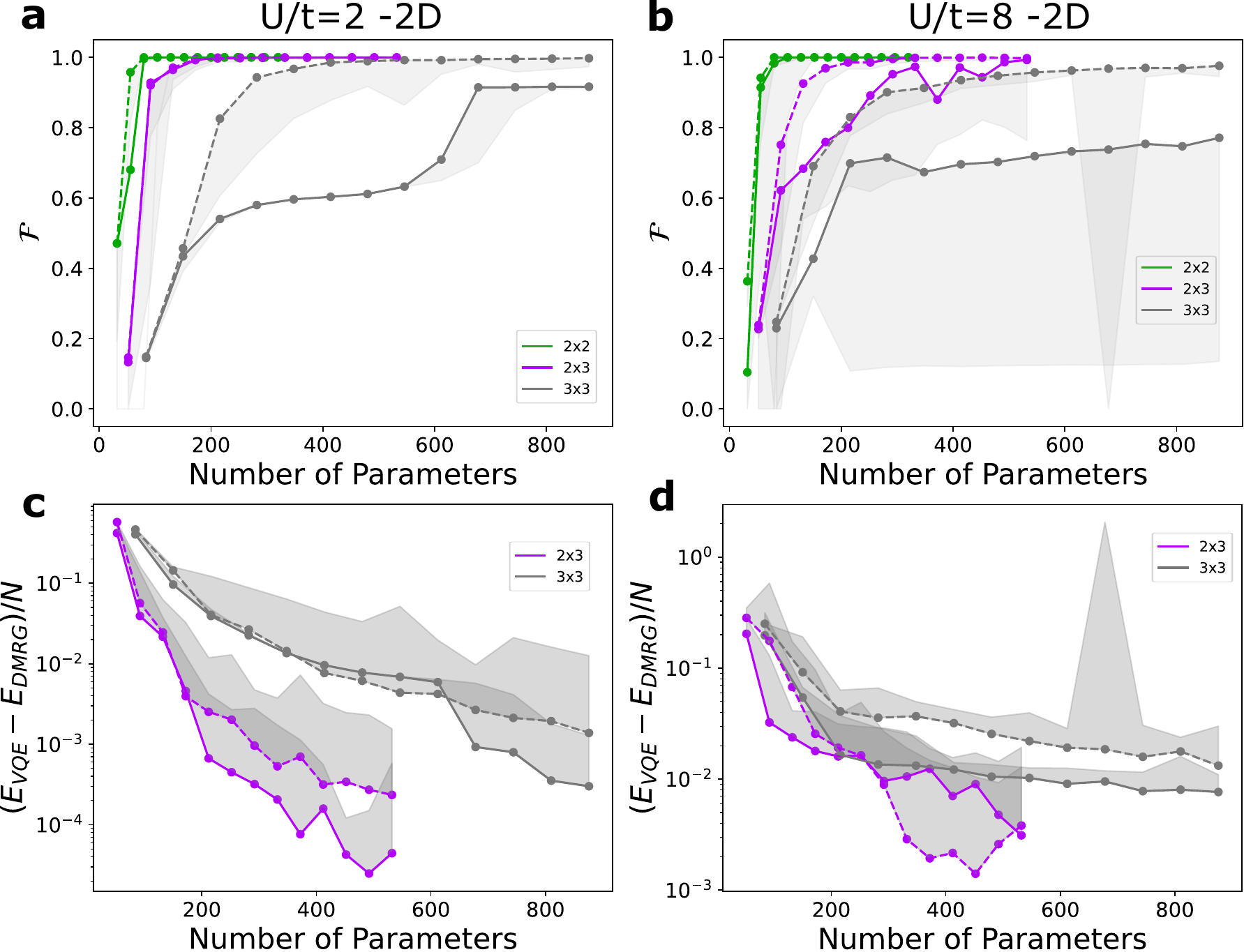}
    \caption{Comparison of the fidelity of the VQE ground state with respect to DMRG, as obtained from an energy-based optimization (solid lines) and an overlap-based optimization (dashed lines), when using the NP ansatz, and for 2D lattices with $U/t=2$ (panel \textbf{a}) and $U/t=8$ (panel \textbf{b}). Comparison of the energy of the VQE ground state with respect to DMRG, as obtained from an energy-based optimization (solid lines) and an overlap-based optimization (dashed lines), when using the NP ansatz, and for 2D lattices with $U/t=2$ (panel \textbf{c}) and $U/t=8$ (panel \textbf{d}). The shaded regions indicate the full range of values found by optimizing ten independent configurations in each case. }
    \label{fig:fidelity}
\end{figure*}

The improved description of long-range correlation functions from an overlap-based VQE approach is because of the higher fidelity of the ground state wavefunctions
produced by this approach. In order to 
systematically benchmark this behavior, 
we visualize in 
Figs.~\ref{fig:fidelity}a and \ref{fig:fidelity}b the fidelity $\mathcal{F}=|\bra{\Psi_{\rm VQE}}\ket{\Psi_{\rm DMRG}}|^2$ of
the VQE ground state from the energy- (solid lines) and overlap-based (dashed-lines) approaches, when using the NP ansatz for 2D lattices, and for $U/t=2$ and $U/t=8$. We see that with the addition of
more parameters to the optimization, both
methods converge towards the limit of $\mathcal{F}=1$, except for the largest system size $3 \times 3$. However, the convergence
of the overlap-based optimization is generally faster, especially for larger
lattices, a feature which is particularly prominent in the case of $U/t=8$, where
strong electronic correlations are present. The fact that the overlap-based
optimization performs better in terms
of yielding higher-fidelity ground states is not surprising, given that
this method is tailored specifically to minimize the loss function of Eq.\,\eqref{eq:loss_overlap}, \emph{i.e.}, to minimize the infidelity of the VQE ground state with respect to the DMRG one. It is worth emphasizing
that the higher fidelity solutions produced
via overlap-based optimization do not always
possess lower energies compared to the lower-fidelity wavefunctions obtained via energy-based minimization, and their energy error
can in fact often be  even somewhat greater than
that from energy-based optimization. The VQE energies obtained with the two approaches are visualized in Figs.~\ref{fig:fidelity}c and \ref{fig:fidelity}d for $U/t=2$ and $U/t=8$ respectively, for $2\times 3$ and $3\times 3$ lattices, 
with all the ground-state energy values of the overlap-based approach given in the Supplemental Material. The higher energies that are produced by the overlap-based optimization are not surprising given the nature of this method. If we consider the solution to be of the form $\ket{\Psi}_{VQE}=\alpha\ket{\Psi}_{GS}+\beta\ket{\chi}$, where $\ket{\Psi}_{GS}$ the true ground state and $\ket{\chi}$ some arbitrary wavefunction, the aim of the overlap-based approach is to produce solutions which maximize the value of $|\alpha|$. However, depending on the nature of the admixture $\beta\ket{\chi}$, for the
energy of which there are no guarantees, the state $\ket{\Psi}$ could assume a range
of energies higher than that achieved through an energy-based minimization. 
Nevertheless, we
see that while the minimum energy
produced by the overlap-based approach
can sometimes be higher than that
of the energy-based approach, in practice the two
results are often within the margin of error of each other. Therefore, given the superior ground state fidelity produced by an overlap-based optimization, this approach can provide
a balanced approach to finding the ground state of the Hubbard model. It also suggests using overlap-optimized wavefunctions as input for further energy minimization, which is feasible even for larger lattices when the DMRG wavefunctions are no longer exact.

\section{Conclusions and outlook}
In this work we have presented a rigorous benchmark of the quantitative features of the solution of the Hubbard model using variational quantum eigensolvers. 
We find that the NP ansatz of Ref.~\cite{Cade2020} yields energies with the smallest error relative to a DMRG reference, compared to the EP and UCC ans\"{a}tze. 
However, even when using the NP ansatz, the error in the ground state energy increases for larger lattices, and ultimately 
plateaus with increasing number of parameters, due to the optimization becoming stuck in local minima. While our exact calculations on classical computers are not limited by barren plateau phenomena, these will make the optimization even more challenging on quantum hardware.
Moreover, we have found that systems with stronger electronic correlations pose an even greater challenge in terms of their
accurate quantitative description, and require more optimization parameters.
At the same time, spatial inhomogeneity in the parameters (\emph{i.e.}, disorder)
does not seem to strongly affect the convergence rate of the results, nor does the addition of nearest-neighbor Coulomb interactions. This is encouraging as we
are moving towards the description of more
complex versions of the Hubbard model, 
representing real materials through schemes
such as \emph{ab initio} downfolding~\cite{Nakamura2021,PhysRevApplied.23.044028}. We expect our results to serve as
an important reference for future applications of VQE for the solution of Hubbard models and strongly-correlated materials, providing an understanding of the accuracy and limitations that can be expected depending on the features of a given system. 

We have also shown, in agreement with previous work, that the qualitative characteristics of the ground state wavefunction of the one-dimensional Hubbard model reproduce the expected magnetic behavior depending on the $U/t$ ratio. We found that while for smaller lattices the VQE solutions obtained via energy minimization can give quantitatively accurate results for spin correlation functions, for larger systems
it struggles to capture long-range correlations. However, a VQE optimization
based on maximizing the overlap to the solution of a classical reference method
succeeds in describing these long-range
phenomena. 
While of course ultimately 
it is the goal of the community to 
utilize quantum computers to obtain the
properties of Hubbard model ground states
for systems beyond the reach of classical
methods, where computing the overlap to
the ``true'' ground state will not be possible, this result still highlights the
potential of an overlap-based optimization
using an \emph{approximate} classical reference state to provide a strong starting point for
energy-based VQE.
The overlap-based optimization results also highlight that the NP ansatz is expressive enough to capture features like long-range correlations. 

Moving forward, it will be important to extend the applicability of current approaches to the solution of Hubbard models with multiple electronic bands, as these multi-band models describe several physical phenomena of relevance to applications, including superconductivity and excitonic insulating behavior~\cite{PhysRevB.85.165135,PhysRevB.89.115134,PhysRevB.98.075117,PhysRevResearch.1.032046}. 
Given the challenge with systematically converging to the global energy minimum with current variational approaches as demonstrated here, even in the one-band case and particularly for the strongly correlated regime, it will be important to develop flexible approaches to achieve this.
One possibility is to explore other ansatz generation schemes, such as adaptive ans\"atze~\cite{Grimsley2019, claudinoBenchmarkingAdaptiveVariational2020, tangQubitADAPTVQEAdaptiveAlgorithm2021, gomesAdaptiveVariationalQuantum2021, mukherjeeComparativeStudyAdaptive2023, vandykeScalingAdaptiveQuantum2023} or problem-inspired ans\"atze tailored for the strongly-correlated regime. Alternatively, optimized variational wavefunctions that have converged to local minima can be used as initial states of a subsequent quantum subspace expansion, which is known to converge faster for a larger overlap between the initial state and the true ground state~\cite{motta2023subspacemethodselectronicstructure,Shen2023realtimekrylov,yu2025quantumcentricalgorithmsamplebasedkrylov,10.1063/5.0217294}. 

\section{Acknowledgments}
 The authors acknowledge useful discussions with Bryan Clark and Erik Gustafson. 
This material is based upon work supported by the U.S. Department of Energy, Office of Science, National Quantum Information Science Research Centers, Superconducting Quantum Materials and Systems Center (SQMS) under contract No. DE-AC02-07CH11359.  We are grateful for support from NASA Ames Research Center. 
Ames National Laboratory is operated for the U.S. Department of Energy by Iowa State University under Contract No. DE-AC02-07CH11358. 
This research used resources of the National Energy Research
Scientific Computing Center, a DOE Office of Science User Facility supported by the Office of Science of the U.S. Department of Energy under Contract No. DE-AC02-05CH11231 using NERSC awards HEP-ERCAP0029167 and DDR-ERCAP0029710.

\textbf{}
\bibliographystyle{quantum}
\bibliography{references}

\end{document}